# Rapid COVID-19 Risk Screening by Eye-region Manifestations

## Authors

Yanwei Fu[1*], Lei Zhao[8*], Haojie Zheng[2*], Qiang Sun[3*], Li Yang[2*], Hong Li[4*], Jiao Xie[8*], Xiangyang Xue[1,5], Feng Li[6], Yuan Li[2], Wei Wang[2], Yantao Pei[2], Jianmin Wang[2], Xiuqi Wu[2], Yanhua Zheng[2], Hongxia Tian[2], Mengwei Gu[1,7,#]

## Affiliations

1. School of Data Science, Fudan University, Shanghai, 200433, China
2. The Fifth Hospital of Shijiazhuang, Hebei Medical University, Shijiazhuang, China
3. Academy for Engineering & Technology, Fudan University, Shanghai, 200433, China
4. Medical Examination Center, Hubei Provincial Hospital of Traditional Chinese Medicine, Wuhan, 430061, China
5. School of Computer Science, Fudan University, Shanghai, 200433, China
6. department of respirology, Shanghai Public Health Clinical Center, Fudan University, Shanghai, 201508, China
7. Aimomics (Shanghai) Intelligent Technology Co., Ltd., Shanghai, 200433, China
8. Department of Infectious Diseases, Union Hospital, Tongji Medical College, Huazhong University of Science and Technology, Wuhan, 430022. China.

## Abstract

The Coronavirus disease 2019 (COVID-19) has affected several million people since 2019. Despite various vaccines of COVID-19 protect million people in many countries, the worldwide situations of more the asymptomatic and mutated strain discovered are urging the more sensitive COVID-19 testing in this turnaround time. Unfortunately, it is still nontrivial to develop a new fast COVID-19 screening method with the easier access and lower cost, due to the technical and cost limitations of the current testing methods in the medical resource-poor districts. On the other hand, there are more and more ocular manifestations that have been reported in the COVID-19 patients as growing clinical evidence[1]. This inspired this project. We have conducted the joint clinical research since January 2021 at the ShiJiaZhuang City, Heibei province, China, which approved by the ethics committee of The fifth hospital of ShiJiaZhuang of Hebei Medical University. We undertake several blind tests of COVID-19 patients by Union Hospital, Tongji Medical College, Huazhong University of Science and Technology, Wuhan, China. Meantime as an important part of the ongoing globally COVID-19 eye test program by AIMOMICS since February 2020, we propose a new fast screening method of analyzing the eye-region images, captured by common CCD and CMOS cameras. This could reliably make a rapid risk screening of COVID-19 with the sustainable stable high performance in different countries and races. For this clinical trial in ShiJiaZhuang, we compare and analyze 1194 eye-region images of 115 patients, including 66 COVID-19 positive patients, 44 rehabilitation patients (nucleic acid changed from positive to negative), 5 liver patients, as well as 117 healthy people. Remarkably, we consistently achieved very high testing results (> 0.94) in terms of both sensitivity and specificity in our blind test of COVID-19 patients. This confirms the viability of the COVID-19 fast screening by the eye-region manifestations. Particularly and impressively, the results have the similar conclusion as the other clinical trials of the globally COVID-19 eye test program[1]. Hopefully, this series of ongoing globally COVID-19 eye test study, and potential rapid solution of fully self-performed COVID risk screening method, can be inspiring and helpful to more researchers in the world soon. Our model for COVID-19 rapid prescreening have the merits of the lower cost, fully self-performed, non-invasive, importantly real-time, and thus enables the continuous health surveillance. We further implement it as the open accessible APIs, and provide public service to the world. Our pilot experiments show that our model is ready to be usable to all kinds of surveillance scenarios, such as infrared temperature measurement device at airports and stations, or directly pushing to the target people groups smartphones as a packaged application.

## Introduction

In December 2019, novel coronavirus disease 2019 (COVID-19) broke out globally. The pathogenic virus was named severe acute respiratory syndrome coronavirus 2 (SARS-CoV-2). As of June 2, 2021, there are about 170 million confirmed cases and 3.7 million deaths worldwide, and the World Health Organization (WHO) declared COVID-19 a public health emergency of international concern[2, 3]. SARS-CoV-2 belongs to the family of Coronaviridae, which is a large family of positive-stranded single-stranded ribonucleic acids with envelopes. Previous studies confirmed that there are six coronaviruses can infect humans, including human coronavirus 229E (HCoV-229E), HCoV-OC43, HCoV-NL63, HCoV-HKU1, Middle East respiratory syndrome coronavirus (MERS-CoV), SARS-CoV[4], SARS-CoV-2 is the seventh member of the coronavirus family that can infect humans[5]. When infected with SARS-CoV-2, the most common symptoms are fever, dry cough, and fatigue, the less common symptoms are expectoration, diarrhea, headache, hemoptysis, chest pain, anorexia, myalgia, chills, nausea and vomiting, and so on[6, 7]. In addition, in small cohort studies and case reports dysosmia and taste disorders like anosmia, phantosmia, parosmia ageusia, dysgeusia also have been reported[8-10]. But it is impossible to make an accurate diagnosis of COVID-19 according to the patient's clinical symptoms. At present, imaging, nucleic acid, and serum antibody are the most common methods used for diagnostic[11], but limited to laboratory or hospital environments, demanding expert-level operations. Thus, patients must be tested in the hospital, and take some waiting time. All these points, greatly limit these state-of-the-art detection methods deployed at large-scale cities, and enabling real-time patient tracking. With the further development of the research, many studies have reported that COVID-19 patients can be combined with different ocular manifestations, such as hyperemia, increased secretion, epiphora, edema, follicular conjunctivitis, scleritis, photophobia, foreign body sensation, itchiness, and so on[11].

Recently, deep learning based classification networks have been widely used to support the disease diagnosis and management[16, 17]. Inspired by these points, we propose a rapid COVID-19 risk screening and diagnosis model with the deep learning method based on eye-region images, captured by normal CCD or CMOS camera and cellphone. We propose the classification model with the corresponding eye images as the input to identify the COVID-19 as a binary classification task. We trained the models by the data from ShiJianZhuang and

Shanghai. The performance was measured at the Area Under receiver-operating-characteristic Curve (AUC), sensitivity, specificity, and accuracy, and F1.

After more than three months of study and clinical trials, we found that the confirmed cases of COVID-19 present the consistent eye pathological symbols, including the asymptomatic. This study included 20 subjects in the pre-test dataset (train and validation) and 212 subjects in the blind test dataset. The subject-level performance of COVID-19 prescreening model in the combined study have achieved an AUC of 0.940(95% CI, 0.888-0.992), 0.990 (95% CI, 0.972-1.000) and 0.975 (95% CI, 0.940-1.000) with respect to the first, second and third blind tests.

As part of the ongoing globally COVID-19 eye test program by AIMOMICS since February 2020, collaborated research worldwide and across racial boundaries has been set up, to call for a more massive database and further refinement to improve the model performance. In the past and ongoing globally registered clinical trials, we show that this model can successfully classify COVID-19 patients from healthy persons, pulmonary patients except for COVID-19 (e.g., pulmonary fungal infection, bronchopneumonia, chronic obstructive pulmonary disease, and lung cancer), liver patients, diabetes patients and ocular patients. The experimental results reveal that patients with COVID-19 have different ocular features from others, which can be used to differentiate them from the public.

The convenient method of eye-region image diagnosis can help the disease control researchers to fully understand the prevalence and pathogenicity of the virus in different ages, time, region, climate, environment, occupation, and population with basic diseases, and guide effective prevention and control measures against COVID-19. Meantime, our model for COVID-19 rapid prescreening have the merits of the lower cost, fully self-performed, non-invasive, real-time results, and thus enables the continuous surveillance. We further implement it as the open accessible APIs, and provide public service to the world. Essentially, our model is also open to other form of embedding for fast screening of COVID-19, such as all kind of infrared temperature measurement device at airports and stations, or directly push to the target people groups' smartphones as a packaged application. We believe a system implementing such an algorithm should assist the large-scale rapid large-scale screening and real-time follow-up, that can be inspiring and helpful for encouraging more researches on this topic.

**Keywords:**

COVID-19, eye-region image, symptom classification, deep-learning, rapid screening,

## Dataset

The data for the study comes from the following three resources,

- COVID-19 patients: all COVID-19 cases were acquired from January to Marth, 2021 by the Fifth Hospital of Shijiazhuang, Hebei Medical University, Shijiazhuang, China. All patients were diagnosed according to the diagnostic criteria of the National Health Commission of China and confirmed by RT-PCR detection of viral nucleic acids. the photos as well as the patient's information are collected. Meanwhile, we obtained the epidemiological, medical history, clinical characteristics, laboratory tests and treatment history of the enrolled patients from electronic medical records and nursing records. Especially, all patients were given chest X-rays or CT. Some of the data needs to be supplemented and confirmed, and we obtain the data through direct communication with the doctor at the bedside, the details of demographics, basic characteristics, clinical characteristics and outcomes of the collected COVID-19 patients are summaries in Table. All patients were taken images according to our data-collection guidelines by two smartphones.

- The rehabilitation patients (nucleic acid changed from positive to negative): all rehabilitation cases were acquired from February to Marth, 2021 by the Fifth Hospital of Shijiazhuang, Hebei Medical University, Shijiazhuang, China. These rehabilitation patients have clear examination reports. we balanced the distribution of age and gender. All cases were taken images according to our data-collection guidelines by four smartphones.

- The liver patients and healthy peoples: all cases as control group were obtained from Marth to April 2021 by Medical Examination Center, Hubei Provincial Hospital of Traditional Chinese Medicine, Wuhan. These liver patients and healthy peoples have clear examination reports. we balanced the distribution of age and gender. All volunteers were taken images according to our data-collection guidelines by three smartphones.

- The data included 1194 photographs (pre-test 101, blind test 1093) from 232 participants (pre-test 20, test 212). We randomly merge the pre-test dataset into the previous work for model updating. The blind test is divided into 3 tests.

- In the pre-test dataset, 20 COVID-19 patients were included, the blind test dataset comprises 66 COVID-19 patients and 166 control group participants.

The demographic and clinical characteristics of COVID-19 patients were shown in **Table 1**.

Table 1. Summary of Developement (Pre-test:Training/Validation), and Testing Datasets. In blind test the detail of control group subject is denoted by 'total(HL/RH/LV)'.

| Dataset | Info | Pre-test | | Test | | | Total |
|---|---|---|---|---|---|---|---|
| | | Hebei Train | Hebei Val | Test1 | Test 2 | Test 3 | Total |
| Data aquisition | date | Jan 1-23 2021 | Jan 1-23 2021 | Feb 1-Apr 30 2021 | Feb 1-Apr 30 2021 | Feb 1-Apr 30 2021 | |
| COVID-19 | subject | 14 | 6 | 16 | 14 | 16 | 66 |
| | images | 68 | 33 | 85 | 85 | 93 | 364 |
| Age | Age | 20-45 | 20-45 | 20-45 | 20-45 | 20-45 | |
| Sex | Male | 4 | 1 | 6 | 4 | 6 | 21 |
| | Female | 10 | 5 | 10 | 10 | 10 | 45 |
| Control Group | subject | 0 | 0 | 53(39/14) | 55(39/16) | 58(39/14/5) | 166 |
| | images | 0 | 0 | 265 | 275 | 290 | 830 |
| Age | Age | | | 21-63 | 21-62 | 21-57 | |
| Sex | Male | | | 24 | 14 | 26 | 64 |
| | Female | | | 29 | 41 | 32 | 102 |
| Total | subject | 14 | 6 | 69 | 69 | 74 | 232 |
| | images | 68 | 33 | 350 | 360 | 383 | 1194 |

In the study there are 2 groups: the COVID-19 group and the control group. The control group comprises of three subgroups: **RH** (rehabilitation patients -nucleic acid changed from positive to negative), **LV** (liver patients) and **HL** (health volunteers). The **CV** (COVID-19 patients) were confirmed by the RT-PCR detection for viral nucleic acids. According to the eighth version guideline published by the National Health Commission of China, asymptomatic/mild COVID-19 carriers refer to individuals with viral nucleic acid test positive, and no clinical symptoms and no signs of pneumonia in chest imaging. The patients with **RH** (rehabilitation patients -nucleic acid changed from positive to negative). The patients with **LV** (liver patients) were diagnosed as hepatopathy. **HL** (the health volunteers) were collected from individuals who had taken physical examination, no obviously abnormal results were demonstrated and no contact history with COVID-19 patients. All the subjects were tested for the COVID-19, and no participants showed viral nucleic acids positive during the following days, except for COVID-19 patients. No death events were observed in this study.

For each participant, 3-10 photographs of the eye-region were captured by using the common CCD and CMOS cameras, assisted by doctors or healthcare workers. The same shooting plain mode and parameters were used, and avoid shooting filters. The photos were captured in a good lighting condition, and not in dark or red background. The image resolution is at least 1900x500 96dpi. The average time for taking a set of 5 eye photos is around 1 minute. In the re-examination step, all the data that cannot reveal the details of the eyes are discarded.

The study was conducted in accordance with the principles of the Declaration of Helsinki. All participants were provided with the written informed consent at the time of recruitment. And this study was approved by the ethics committee of the fifth hospital of ShiJiaZhuang of Hebei Medical University (approval No.: 2021005), and was approved by ClinicalTrials.gov PRS (ClinicalTrials.gov ID: NCT04907981).

## Methodology

First, we observed the pictures with an unaided eye, and found that among the 212 patients who were included(20 patients with pre-test were not counted), there were 21 patients (9.9%) combined with ocular manifestations. In the COVID-19 group, 17 patients (36.9%) showed ocular manifestations(Figure 1), including conjunctival congestion(N=13), increased secretion(N=4), conjunctival hemorrhage(N=1), and ptosis(N=1). While only 4 cases (9.1%) showed conjunctival congestion(N=4) and ptosis(N=2) in the rehabilitation patients. At the same time, no ocular manifestation was observed in liver disease and healthy control group.

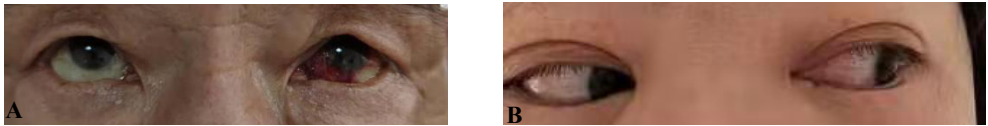

**Figure 1 Cases of conjunctival hemorrhage(A) and conjunctival congestion (B)**

Then, in order to evaluate the study, we develop a model from the previous work and the Hubei pre-test data, using eye-region photos on the dataset. The blind test group includes COVID-19 positive patients, and control group as rehabilitation patients (nucleic acid changed from positive to negative), liver patients, healthy people. As shown in Figure 2, the experiment is divided into 3 blind tests. During the blind test on each group, the data from the previous blind test is used as extra training data. As for the COVID-19 category, the average sensitivity target is no less than 80%, which could demonstrate that the efficacy of our method to distinguish COVID-19 patients.

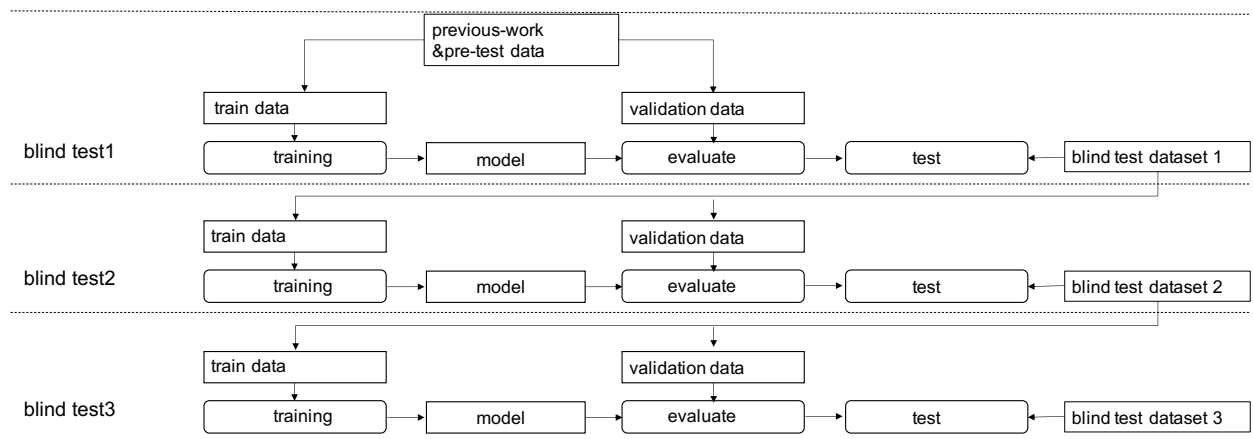

**Figure 2. Study design and workflow of this study.**

**In the first blind test, the pre-test data is combined to improve the model from previous-work. During the second and third blind, the previous blind test data is used as extra data for training and validation.**

The pipeline of our proposed method is illustrated in Figure 3, which is composed of two steps, the image-level classification, and the subject-level classification. The classification network is built by the deep learning neuron networks to study the characteristics of the cropped eye regions in a latent space. Specifically, we define the subject as a set of photos taken by the same people. The classification result of the same person in might be different due to the noise in data and a proper vote strategy can enhance the robustness of the model. The results of different images of the same subject are merged by a special vote strategy. Finally, the subject-level risk assessment of COVID-19 is generated.

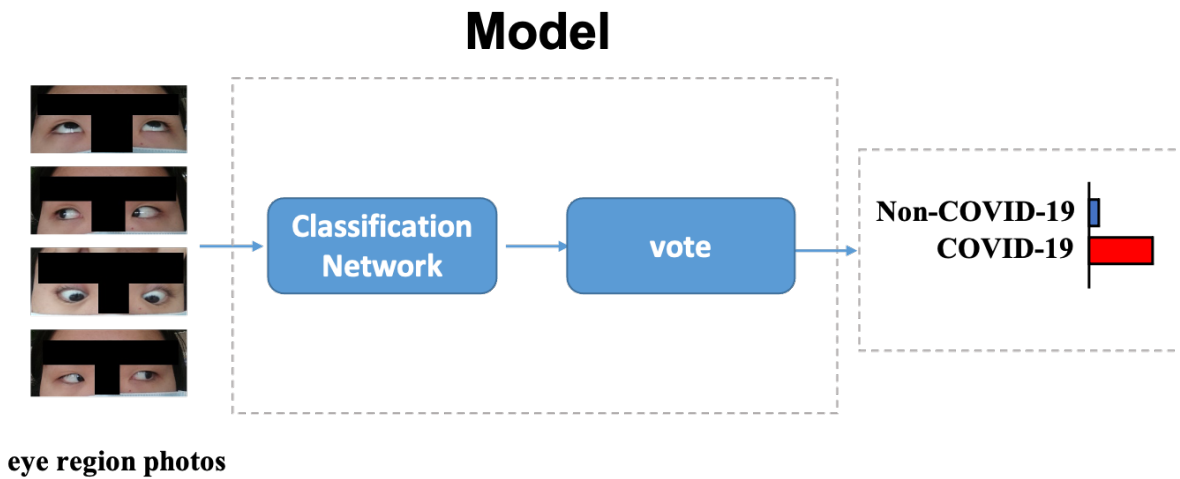

**Figure 3. Illustration of the framework.**

To measure the performance of the binary classification network, we define a set of metrics, such as the area under the receiver-operating-characteristic curve (AUC), sensitivity, specificity, and accuracy. Bootstrapping strategy with 1000 replicates was used to estimate 95% confidence intervals of the metrics, with an image-level resampling unit. Specifically, the receiver operating characteristic curves (ROCs) were used to illustrate the performance in screening COVID-19 disease.

The classification network has achieved an AUC of 0.940(95% CI, 0.888-0.992), 0.990(95% CI, 0.972-1.000) and 0.975(95% CI, 0.940-1.000) for the first, second, and third blind tests on the subject-level classification. The metrics including AUC, sensitivity, specificity, accuracy, and F1 are shown in Table 2. The image-level, max-voting ROC is plotted in Figure 4.

**Table2. Classification Performance of the classification network on the test dataset.**

| Setting/Metric | AUC (95% CI) | Sensitivity (95% CI) | Specificity (95% CI) | ACC (95% CI) | F1 (95% CI) |
|---|---|---|---|---|---|
| Image-Level (blind test 1st) | | | | | |
| COVID-19 vs. Non-COVID-19 | 0.925 (0.898-0.952) | 0.788 (0.701-0.876) | 0.879 (0.839-0.920) | 0.857 (0.820-0.894) | 0.728 (0.656-0.801) |
| Subject-Level (blind test 1st) | | | | | |
| COVID-19 vs. Non-COVID-19 (Max-Voting) | 0.940 (0.888-0.992) | 1.000 (1.000-1.000) | 0.830 (0.729-0.932) | 0.870 (0.789-0.950) | 0.780 (0.630-0.931) |
| Image-Level (blind test 2nd) | | | | | |
| COVID-19 vs. Non-COVID-19 | 0.985 (0.976-0.994) | 0.918 (0.857-0.978) | 0.931 (0.901-0.961) | 0.928 (0.901-0.955) | 0.857 (0.802-0.913) |
| Subject-Level (blind test 2nd) | | | | | |
| COVID-19 vs. Non-COVID-19 (Max-Voting) | 0.990 (0.972-1.000) | 1.000 (1.000-1.000) | 0.855 (0.758-0.951) | 0.884 (0.806-0.962) | 0.778 (0.620-0.935) |
| Image-Level (blind test 3rd) | | | | | |
| COVID-19 vs. Non-COVID-19 | 0.948 (0.917-0.979) | 0.892 (0.828-0.957) | 0.938 (0.911-0.965) | 0.927 (0.901-0.953) | 0.856 (0.802-0.910) |
| Subject-Level (blind test 3rd) | | | | | |
| COVID-19 vs. Non-COVID-19 (Max-Voting) | 0.975 (0.940-1.000) | 0.875 (0.704-1.000) | 0.879 (0.794-0.964) | 0.878 (0.802-0.955) | 0.757 (0.588-0.926) |

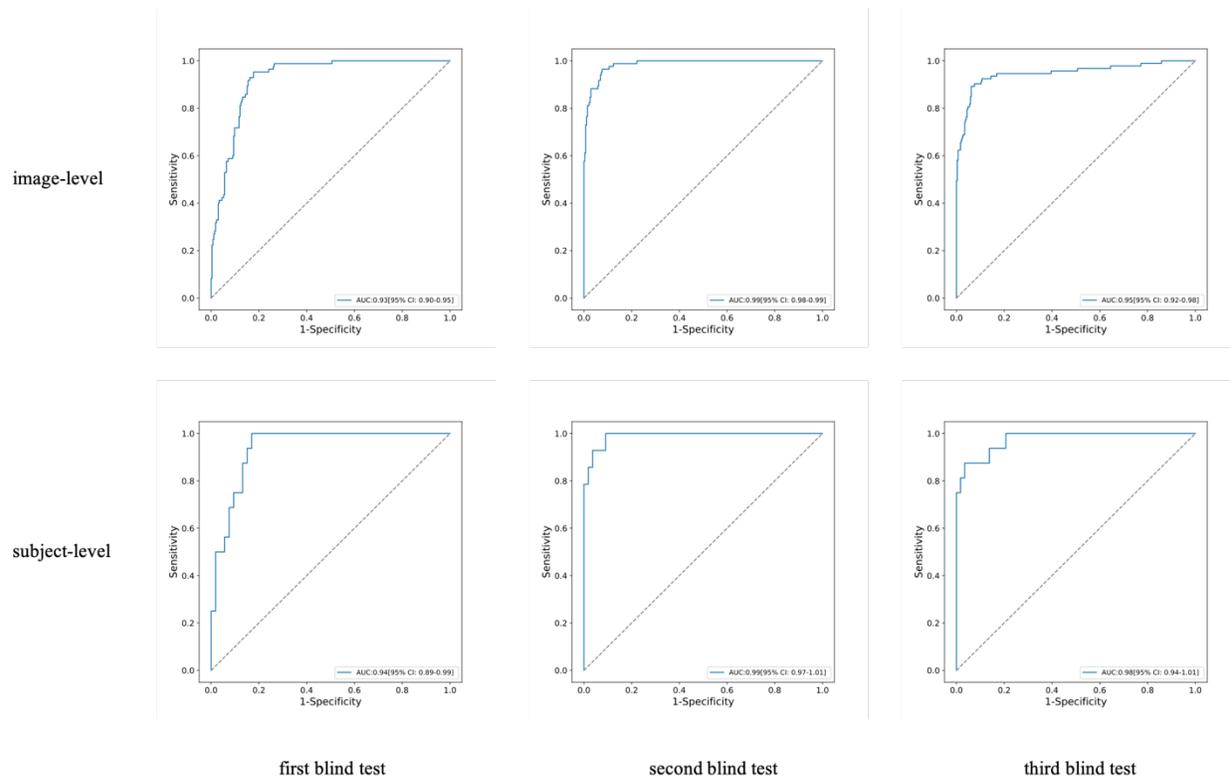

**Figure 4. COVID-19 vs. Non-COVID-19 ROC curves on the three blind tests. The first row is image-level and the second row is subject-level.**

The overall classification system performed well, misjudgment also appeared. In the blind test, the true positive, false positive, false negative and true negative results are summarized in the confusing matrix on both the subject-level and image-level, and the details were shown in **Table 3**.

**Table 3. The confusion matrix of classification result of subjects and images on the test dataset**

|     | GT/Pred | Subject-level(max) | | | | Image-level | | | |
| --- | --- | --- | --- | --- | --- | --- | --- | --- | --- |
|     |     | P | N | P(%) | N(%) | P | N | P(%) | N(%) |
| 1st | P | 16 | 0 | 100 | 0 | 67 | 18 | 78.8 | 21.2 |
|     | N | 9 | 44 | 17 | 83 | 32 | 233 | 12.1 | 87.9 |
| 2nd | P | 14 | 0 | 100 | 0 | 78 | 7 | 91.8 | 8.2 |
|     | N | 8 | 47 | 14.5 | 85.5 | 19 | 256 | 6.9 | 93.1 |
| 3rd | P | 15 | 2 | 88.2 | 11.8 | 83 | 10 | 89.2 | 10.8 |
|     | N | 7 | 51 | 12.1 | 87.9 | 18 | 272 | 6.2 | 93.8 |

To investigate the interpretability of the model, we conduct a visual analysis of the key areas of the model's attention in the classification process. The key areas of the model's attention were converted into heat maps based on gradients and activation maps by GradCAM.

For case study heatmap in **Figure 5**, we found that the attention of heatmaps of Non-COVID-19 and COVID-19 mainly covers the iris, upper and lower eyelid, the inner and outer eye corner.

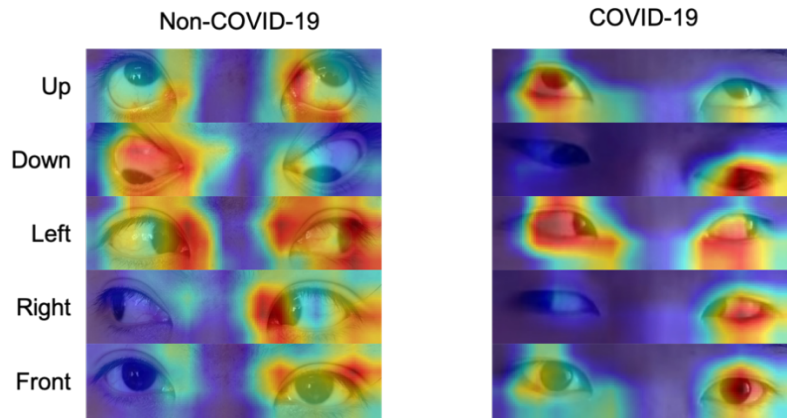

Figure 5. Case heatmap.

## Discussion

COVID-19 is an infectious disease caused by SARS-CoV-2. The patients usually have fever, cough, fatigue, shortness of breath, and loss of smell and taste. Nowadays, some researchers also find that Olfactory and

Gustatory Dysfunction is one of the common symptoms of the COVID-19. Meantime, more ocular manifestations have been reported in the COVID-19 patients as growing clinical evidence[18-20].

Accordingly, abnormally high body temperature has been the initial and most widely used screening method for COVID-19 patients. Meanwhile, the emergence and rising number of asymptomatic COVID-19 patients are posing a threat to body temperature-based screening methods. On one hand, despite the high sensitivity and good specificity of detecting the nucleic acid sequence of pathogens and amplify the signal by an amplification reaction, the long process of RT-PCR may remarkably increase the risk of infection, less ideal for epidemic control. On the other hand, conventional CT tests can show that patients with COVID-19 can have the bilateral ground-glass appearance and pulmonary turbidity, analyzed by radiologists, or deep learning algorithms, however, for some patients, CT imaging results are not obvious or even negative. Particularly, there are many other bottlenecks and difficulties for the COVID-19 diagnosis globally, such as the restriction of pharyngeal swab sampling, the lack of RT-PCR kits, the difficulties in the transportation, the preservation of samples, political factors in some country, and so on. The deep learning based classification network can achieve impressive performance on the eye-region images. And the heatmaps generated by deep learning model have shown that the visual features of different diseases are mainly focused on the eye region, which indicates that the model has learned the key area for disease diagnosis.

Various publications showed the ocular involvement with Coronavirus family[21-23]. In our study, the Ocular symptoms of COVID-19 patients accounted for about 36%, the main manifestation is conjunctivitis, which was consistent with the results of previous studies[15, 24]. SARS-CoV-2 infects the body through the binding of angiotensin converting enzyme 2 (ACE2) receptor[25, 26], it is widely expressed in a variety of cells, primarily in the lungs, kidneys, heart, gastrointestinal tract, and liver[25, 26]. Recently, it has also been reported that ACE2 expressed in conjunctiva[28]. Its expression in conjunctiva may also be related to the occurrence of COVID-19 conjunctivitis.It is worth noting that some of the recovered patients still have ocular manifestations, and there are also some patients are positive in the model test. Studies have shown that the ocular performance of rehabilitated patients is accompanied by a significant increase in the level of inflammatory factors[29, 30], which indicate the eye performance is not caused by the direct invasion and destruction of SARS-CoV-2, but the virus causes a wide range of pro-inflammatory cytokines and chemokine responses in conjunctival and limbal epithelial cells[29, 30], resulting in a proliferation of cytokines, resulting in apoptosis of peripheral corneal epithelial cells and reactive inflammatory injury. Therefore, we believe that longer and more frequent follow-up is necessary for the recovered patients.

Meanwhile, there are some limits to this study. First, the study sample size is small and most COVID-19 patients were collected from East Asia (China). A larger multicenter study covering more patients of diverse race would be necessary to test the performance of the ocular surface feature-based deep learning system. At last, the pathological significance of extracted features from COVID-19 patients should be carefully interpreted and re-verified by ophthalmologists.

## CONCLUSION

Our deep learning method in this clinical trial study, which discriminates over 85% COVID-19 positive patients from the test group, including asymptomatics. Consider that eye exam technology has been used to screen a variety of diseases, such as diabetes and kidney disease. In this paper, we proposed a deep learning model for rapidly risk screening COVID-19 with eye-region images. Different from previous studies, which utilize RT-PCR or CT imaging, the input of our system is the face image or binocular image captured by common CCD and CMOS cameras. Combining with the development of deep learning, it enables the real-time COVID-19 screening from two aspects, sample acquisition, and testing.

We are conducting the large-scale experiments to further validate the effectiveness and efficacy of our algorithm as the tool of the new screening method. On the other hand, due to the privacy policy and the difficulty of data collection in different countries by racial diversity, we have furthered investigate and extended the capability of our algorithm in few-shot learning settings. We have achieved sustainable stable high performance in a series of registered clinical trials at different counties, thanks to our previous works[32, 33] from the computer vision and machine learning communities.

We believe that this study can be inspiring and helpful for encouraging more researches in this direction, and provide effective and rapid assist for clinical risk screening, especially during outbreaks. To conduct further research, we are building a globally COVID-19 eye test free API and Demo links platform which implements our algorithms and helps fast screening COVID-19. Furthermore, this platform could gradually open more than 300 diseases eye tests and joint study programs' invitation, such as virus influenza, diabetes, hepatopathy, etc. through the open accessible APIs or Demo links, the test algorithm could be easily deployed or embedded in the HD camera and any detect accessories, combined into a multi-modal approach including vision and other sensors, continually monitoring the particular diseases control districts were around the public access, transportation hub, population centre, quarantine house, making everyone health care more accessible with lower cost.

So we are looking forward to more scientists, engineers and application developers join us to explore the eye testing and collaborative work together on the critical challenges, ensuring "equality、altruism and ethics

" are the highest principle.For more study information and the access license for joining our the globally COVID-19 eye test program, please contact Dr. Yanwei Fu yanweifu@fudan.edu.cn.

## Data-collection guidelines

For scientific research, we need some eye region data. We promise that the data will not be used for commercial, only for scientific research. The specific requirements are as follows:

1. The eyes in the image need to be clear, a total of five required angles are shown on the left.

2. When taking photos, please do not wear cosmetic products such as contact lenses.

3. When taking photos, please do not use beauty camera mode, or post-production beauty filters, we need the original images.

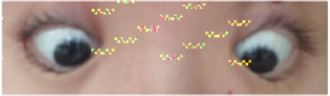

1. 下视 look down

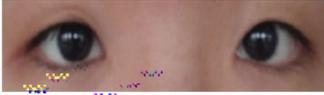

2. 平视 look horizontally

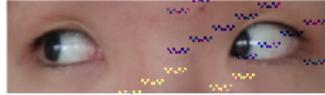

4. 右视 look right

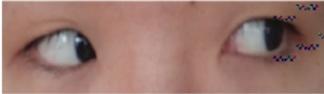

3. 左视 look left

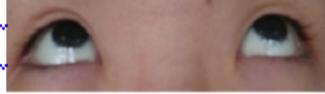

5. 上视 look up

## 1. Sampling equipment and environmental requirements

1) For the above sampling, the same model of mobile phone or shooting equipment must be used to prevent the sampling data domain from being interfered with by the equipment.

2) If the conditions are not available, it is necessary to use the same model of mobile phone to collect all the sampling images at the same time, to maximize the elimination of device data domain interference.

3) The same shooting parameters must be used when shooting, and the beauty, soft light, and other shooting filters must not be used.

4) The image resolution of the eyes is at least: 1900x500 96dpi

5) The shooting environment should be well-lit and bright. It should not shoot in front of dark and red backgrounds. A white background is best.

## 2. Sampling compliance

a) The subject or legal representative has signed an informed consent form

b) Age ≥18 and ≤75 years old (under 18 years old, the guardian shall sign the informed consent form)

c) Meet the diagnostic criteria for COVID-19 infection

## 3. Ethics committee approval

All participants were provided with written informed consent at the time of recruitment. And this study was approved by the ethics committee of The Fifth Hospital of ShiJiaZhuang of Hebei Medical University. The study was in accordance with the Declaration of Helsinki 1964 and its successive amendments.

## 4. Authors' contributions

Dr.Yanwei Fu conceived of the presented idea and gave the eye test conceptualization, conceived, designed and coordinated the study, revised the manuscript; MD PhD Lei Zhao, MD PhD Haojie Zheng, MD PhD Jiao Xie provided major medical guidance, conceived and designed experiments, provided medical guidance and participated in the paper writing, gave model validation; Dr. Qiang Sun performed all of the data analyses and visualization presentation, improve the model, participated in the paper writing; MD PhD Li Yang, MD PhD Yuan Li, MD PhD Wei Wang, MD PhD Yantao Pei, MD PhD Jianmin Wang, MD PhD Xiuqi Wu, MD PhD Yanhua Zheng, MD PhD Hongxia Tian collected the clinical data from the Fifth Hospital of Shijiazhuang, Hebei Medical University, Shijiazhuang, China, provided medical guidance and participated in the paper writing. MD PhD Hong Li collected the clinical data from the Medical Examination Center, Hubei Provincial Hospital of Traditional Chinese Medicine, Wuhan, China, provided medical guidance; MD PhD Feng Li provided key study resources and medical guidance, gave model validation; Dr. Xiangyang Xue provided key study resources, revised the manuscript; Dr. Mengwei Gu gave the eye test conceptualization, conceived, designed and coordinated the study, revised the manuscript and supervised the project

## 5. Conflict of interest statements

All other authors declare no competing interests.

## 6. Acknowledgments

We would like to thank Dr. DaiErHei and Dr. Yu Liu for their kind assistance with this research project.